\documentclass[letterpaper,english,aps,pra,twocolumn,floatfix, showpacs, amsfonts, amssymb]{revtex4-1}
\usepackage{epsfig, graphicx,psfrag,amsmath,amssymb,float}
\usepackage[T1]{fontenc}
\usepackage[latin9]{inputenc}
\usepackage{amsmath}
\usepackage{amssymb}

\usepackage{amscd}
\usepackage{bm}
\usepackage{psfrag}

\usepackage{babel}

\newcommand{\be}{\begin{equation}}
\newcommand{\ee}{\end{equation}}
\newcommand{\bea}{\begin{eqnarray}}
\newcommand{\eea}{\end{eqnarray}}

\begin{document}

\title{Mobile impurities and orthogonality catastrophe in two-dimensional vortex lattices}
\author{M. A. Caracanhas}
\author{R. G. Pereira}
\affiliation{Instituto de F\'{i}sica de S\~ao Carlos, Universidade de S\~ao Paulo, C.P. 369, S\~ao Carlos, SP,  13560-970, Brazil}

\date{\today}
\begin{abstract}
We investigate the properties of a neutral impurity atom  coupled with the Tkachenko modes of a two-dimensional vortex lattice in a Bose-Einstein condensate. In contrast with polarons in homogeneous condensates, the marginal  impurity-boson interaction in the vortex lattice leads to infrared singularities  in perturbation theory and to the breakdown of the quasiparticle picture in the low energy limit.  These infrared singularities are interpreted in terms of a renormalization of the  coupling constant, quasiparticle weight and  effective impurity  mass. The divergence of the effective mass in the low energy limit gives rise to a  power law singularity  in the impurity spectral function and provides an example of an emergent orthogonality catastrophe in a bosonic system.

\end{abstract}


\maketitle

\section{Introduction}
Ultracold quantum gases are by now well established as an alternative platform  for realizing interacting  models originally developed in  condensed matter physics \cite{bloch}. Moreover, the high controllability  of  model parameters     and   system dimensionality  allows one to  explore exotic phases of matter that could never be reached in conventional experiments involving electrons in metals. In particular,  recent techniques for cooling two-component  mixtures   have opened  the way to investigating  polaron physics via the interaction of a low density of ``impurity'' atoms with the environment formed by another majority  species \cite{mathey,StrongCoup,jaksch,pupillo,zwierlein,drevessen3,kohl,zaccanti1,berciu,fukuhara,balewski,zaccanti2}.  The properties of these atomic polarons can be probed using species-selective radio-frequency spectroscopy \cite{T1,shashi}, and the experimental results can be compared quantitatively  with predictions from microscopic many-body theory even in the  strong coupling limit \cite{zaccanti2}.

In ultracold atom systems, there is a diverse  set of collective modes with which the impurity can be dressed to form a polaron. For instance, the interaction  of impurities with  particle-hole pairs  of a Fermi sea has been studied  in Fermi gases with a large population imbalance  \cite{kohl,zwierlein,zaccanti1,nascimbene,parish,schmidt}. By tuning the atomic $s$-wave interaction, one can even   switch between    the regimes of    attractive  and repulsive polarons. In the case of a background formed by bosonic atoms,  polarons are formed when the impurities are dressed by the Bogoliubov phonons of a condensate \cite{jaksch,drevessen3}. Interestingly, the phonon-mediated interaction between impurities can lead to the formation of polaron clusters and ``localization'' in the sense of broadening of the momentum distribution \cite{jaksch}.

In Ref. \cite{PRL111} we proposed a novel  addition to the polaron family which we referred to as the  ``Tkachenko polaron''. The latter originates  when an impurity  atom is immersed in a  two-dimensional (2D) vortex lattice  \cite{Science,cornell} formed by ultracold bosons in the mean field quantum Hall regime \cite{PRL87,muller}.  The vortex lattice is distinguished by the existence of so-called Tkachenko modes with parabolic dispersion
\cite{sonin,baym,sinova,shlyapnikov,shlyapnikov2}, as opposed to the usual linear dispersing phonons of a 2D crystal. Using perturbation theory, we showed that at weak coupling the polaron spectral function has a Lorentzian lineshape with a decay rate linearly proportional to the polaron energy.  On the other hand, a renormalization group (RG) analysis of the effective  field theory in the continuum limit reveals that  the interaction between the impurity and the bosonic modes with quadratic dispersion in 2D is marginally relevant. This implies that the effective coupling grows as the energy decreases, leading to an anomalous broadening of the spectral function  and the breakdown of the quasiparticle picture   in the long wavelength limit.

In this work we wish to clarify the properties of the low energy limit of Tkachenko polarons. We shall show   that the effective Hamiltonian flows  under the RG to a line of fixed points in which the impurity mass diverges  while the  impurity-boson coupling  is enhanced but remains finite. In this limit the heavy impurity is dressed by a diverging number of low-energy modes,  a characteristic feature of the orthogonality catastrophe (OC) phenomenon  \cite{mahan,noziere,noziere2,noziere3}. The signature of this phenomenon on the impurity spectral function is the development of an approximate power-law singularity at low energies. There are two unique aspects about the OC   in our model. First, it arises in an impurity model with a background of \emph{bosonic} excitations, whereas the usual OC (which has also been studied in the context of cold atoms \cite{T2,PRX,demler,Ap2}) occurs in fermionic systems. Second, while the usual OC requires a localized impurity, in our case we can start with a mobile impurity  ({\it i.e.} with a finite mass), and the localization   emerges asymptotically   as the effective impurity mass diverges  in the low energy limit. We can then speak of \emph{self-trapping} in the sense that the time scales over which the impurity  moves through the lattice become anomalously large.  Both  of these aspects  stem from  the peculiar dispersion of Tkachenko modes, which exhibit a finite density of states in the low energy limit, leading to infrared singularities in perturbation theory. We note that similar infrared singularities occur in one-dimensional systems \cite{kantian}.

The paper is organized as follows. In Section II we review  results for the Tkachenko polaron model in the weak coupling regime \cite{PRL111}. The RG flow of the parameters in the effective model   is calculated in Section III. Next, in Section IV, we analyze the low energy fixed points  by applying a canonical transformation to extract the power law singularity characteristic of the OC. This section presents our main results concerning the   spectral function for the mobile impurity in  the low energy limit. In Section V we point out connections to existing experiments and the viability of detecting our predictions. Finally, we summarize our results in Section VI.

\section{Weak coupling regime of the light Tkachenko polaron}

\subsection{Continuum model}

We consider a two-component boson mixture with a large population imbalance between species $A$ (majority atoms) and $B$ (impurity atoms). We assume that both species  occupy the ground state of a strongly confining  potential in the $z$ direction,  so the system is effectively  2D. Within the $xy$ plane the bosons are confined   by a weaker harmonic trap $V_{ext}(\bf{r})$.

We would like to  induce a vortex lattice in the majority species  while keeping the impurities with nearly free dispersion. This is not possible in a rotating trap since the Hamiltonian in the rotating frame  contains an effective magnetic field that couples to both species \cite{cooper}, leading to the undesirable effect of Landau quantization of the impurity energy levels. For this reason, we consider instead an artificial vector potential ${\bf A}(\bf{r})$  \cite{dalibard2,lin} that corresponds to an effective uniform magnetic field  in the laboratory frame and couples selectively to $A$ atoms.   The essential idea to implement species-specific artificial gauge fields for neutral atoms is to produce Berry phases by combining the internal atomic structure with carefully engineered optical potentials, for instance via spatial gradients of detuning or Rabi frequency \cite{spielman}. The interacting Hamiltonian in the presence of the gauge fields is $H = H_{A}+H_{B}+H_{int} $,  with \begin{eqnarray} \nonumber
H_A& =& \int d^2r \Bigg[ \hat\psi_A^\dagger\frac{(-i\hbar\nabla-{\bf A})^2}{2m_A}\hat\psi_A^{\phantom\dagger} +V_{ext}(\mathbf{r})\,\hat\psi_A^\dagger\hat\psi_A  \\  \nonumber  &+&  \frac{g_{A}}{2}(\hat\psi_A^{\dagger}\hat\psi_A^{\phantom\dagger})^2\Bigg],\\
\nonumber H_B &=& \int d^2r \left[\hat\psi_B^\dagger\frac{(-i\hbar\nabla)^2}{2m_B}\hat\psi_B^{\phantom\dagger}+V_{ext}(\mathbf{r})\,\hat\psi_B^\dagger\hat\psi_B\right.\\
&&\nonumber\left. +\frac{g_{B}}{2}(\hat\psi_B^{\dagger}\hat\psi_B^{\phantom\dagger})^2\right], \label{Hbec}\\
H_{int} &=&g_{AB}\int d^2r\,\hat\psi_A^{\dagger}\hat\psi_A^{\phantom\dagger}\hat\psi_B^{\dagger}\hat\psi_B^{\phantom\dagger}.
\end{eqnarray}
Here each species   is described by a creation (annihilation) operator $\hat{\psi}_{i}(\mathbf{r}) \; [\hat{\psi}_{i}^{\dag}(\mathbf r)]$, with $i=A,B$. The intra-species repulsive contact interactions for the 2D system are  given   by $g_{i}=2\sqrt{2\pi}\hbar^2a_{i}/m_{i}l_{0}$, and the inter-species interaction is $g_{AB}=\sqrt{2\pi}\,\hbar^2\,a_{AB}/\mu  l_{0}$, where $\mu=m_{i} m_{i}/(m_{i}+m_{j})$ is the reduced mass, $a_A,a_B,a_{AB}$ are the corresponding three-dimensional $s$-wave scattering lengths, and $l_{0} =\sqrt{\hbar/m_{A}\omega_{0}}$  is the  axial oscillator length for a harmonic trapping potential with frequency $\omega_0$ in the $z$ direction.

The vorticity of the $A$ subsystem is characterized by the magnetic length $l = \sqrt{\hbar/\mathcal{B}}$, with $\mathcal{B}=|\nabla\times \mathbf{A}|=$ const., or by the cyclotron frequency $\Omega = \mathcal{B}/m_A$.  At critical  vorticity, {\it i.e.} when the  oscillator length of $V_{ext} $ matches  the magnetic length, the residual confining potential for $A$ atoms in the $xy$ plane  vanishes and the system is effectively in an infinite plane geometry \cite{shlyapnikov2}. Let $n_A=N_A/\mathcal{S}$ denote the average 2D density for $N_A$ atoms distributed over an area $\mathcal{S}$.  The 2D density of vortices is  $n_V = N_V/\mathcal{S}= (\pi l^2)^{-1} $, and the filling factor is $\nu = N_A/N_V=n_A \pi l^2$ . In  the mean-field quantum Hall regime \cite{PRL87,muller}  $g_{A}n_A\ll    \hbar   \Omega$,  a good starting point  is to consider that  all $A$ atoms occupy the same macroscopic quantum state  given by a linear superposition of  lowest Landau level states. The mean field state  that minimizes the energy is   the Abrikosov vortex lattice state  $\psi_{A}(\mathbf{r}) = \sqrt{n_A}\varphi_A(\mathbf{r})$, where $\varphi_A(\mathbf{r})= (2\varsigma)^{1/4} \;\vartheta_{1}(\sqrt{\pi \varsigma} z,\rho) \; e^{z^{2}/2}\;e^{-|z|^{2}/2}$  is a normalized wavefunction involving the Jacobi theta function $\vartheta_{1}$ with  parameters $z=(x+iy)/l$, $u = -1/2$, $\varsigma= \sqrt{3}/2$,  $\tau = u +i\varsigma$,  and $ \rho =\exp(i\pi\tau)$. One can check that the density profile $|\varphi_A(\mathbf{r})|^2$ corresponds to  a triangular vortex array  \cite{shlyapnikov2}.

Before advancing with the model, a few remarks about the stability of the vortex lattice are in order. In the mean field regime, the number of vortices $N_V$ is well below the number of $A$ atoms. In other words, the filling factor is large, $\nu \gg 1$;  experimental values  of  $\nu \sim 500$ have been reported \cite{cornell,cooper}. It is known that even at $T=0$ the vortex lattice state lacks long-range phase coherence \cite{sinova}. However, the crystalline density profile is stable against quantum melting  for  filling factors above a critical value $\nu_c\sim 6$ \cite{cooper2}. Below this critical value, incompressible quantum Hall phases of bosons have been predicted  \cite{cooper}. Here  we are interested in the regime $N_B\ll N_V\ll N_A$, where we can safely assume that the vortex lattice is stable against its own quantum fluctuations as well as against perturbations induced by coupling to  dilute $B$ atoms.

The  excitation spectrum of the vortex lattice  can be obtained by expanding  $H_A$   about  the mean field  solution \cite{sinova,baym,shlyapnikov}.  The spectrum contains one gapped ``inertial'' mode and one gapless mode --- the Tkachenko mode --- with parabolic dispersion in the low energy limit. The latter corresponds to the Goldstone boson expected from spontaneous breaking of translational and rotational symmetries in the vortex lattice state. The  parabolic dispersion may seem unusual, but is consistent with the counting of Goldstone modes for non-relativistic systems \cite{nielsen,watanabe}. In the mean field regime the inertial mode gap (of order $\hbar\Omega$)  is large and the lattice dynamics  is dominated by the gapless Tkachenko mode. Following \cite{shlyapnikov}, we expand the field operator for $A$ atoms about the mean field state in the form   $\hat{\psi}_{A} = \psi_{A} + \delta\hat{\psi}_{A}$, with   \begin{equation}
\delta\hat\psi_A(\mathbf r) = \frac1{\sqrt{ \mathcal S}} \sum_{\mathbf q\in \textrm{BZ}}\left[u_{\mathbf q}(\mathbf r)a_{\mathbf q}^{\phantom\dagger}-v_{\mathbf q}(\mathbf r)a_{\mathbf q}^{\dagger}\right].\label{deltapsiA}
\end{equation}
Here $a_{\mathbf q}$ is the annihilation operator for the Tkachenko mode with wave vector $\mathbf q$ defined in the Brillouin zone of the triangular lattice and  $u_{\mathbf q}(\mathbf r)$, $v_{\mathbf q}(\mathbf r)$ are solutions of the projected Bogoliubov-de Gennes equations. For $q\ll l^{-1}$, the dispersion relation is $\hbar \omega_{\mathbf q}\approx \hbar^2q^2/2M$, with the effective  mass $M\sim (\hbar\Omega/n_A g_A)m_A\gg m_A$.

The Tkachenko polaron  is defined as the problem of a single $B$ atom propagating  in the background of a vortex lattice state \cite{PRL111}.
The first difference from a homogeneous condensate appears to zeroth order in the fluctuations $\delta\hat\psi_A$, in the form of a static lattice potential  obtained  by substituting the mean field solution for $\psi_A$ in $H_{int}$ in Eq. (\ref{Hbec}). However, in the limit of weak interspecies interaction $n_Ag_{AB}\ll \hbar \Omega$ and small  momenta $q\ll l^{-1}$, the effective  impurity mass $m_B$   is only  weakly renormalized by the shallow lattice potential. We assume $m_B\sim m_A\ll M$, so that the   impurity is light compared to the Tkachenko boson. Hereafter we set $m_B=m$ to lighten the notation.

The  term generated by $H_{int}$ to first order in the fluctuation $\delta\hat\psi_A$ is an effective ``impurity-phonon'' interaction, with  phonons replaced by   Tkachenko modes. In the continuum, large-polaron limit $k,q\ll l^{-1}$, we obtain \cite{PRL111} \begin{eqnarray} H_{imp-ph} \approx\frac{\lambda}{\sqrt{\mathcal{S}}}\sum_{\textbf{k},\textbf{q}}   |\mathbf{q}|\hat{b}_{\textbf{k+q}}^{\dagger}\,\hat{b}^{\phantom\dagger}_{\textbf{k}}\,(\hat{a}_{\textbf{q}}+\hat{a}_{-\textbf{q}}^{\dagger}),\end{eqnarray}
where $\hat b_{\mathbf{k}}$ is the annihilation operator for impurities in states with momentum $\mathbf{k}$ and  $\lambda\sim \sqrt{\nu}g_{AB}$ is  the impurity-boson coupling constant. Note that  $\lambda$ is enhanced by the large filling factor $\nu\gg1$.  We finally obtain the 2D Tkachenko polaron model  \cite{PRL111} \begin{eqnarray} \nonumber \label{eqH}
H &=&H_{ph}+H_{imp}+H_{imp-ph}\nonumber\\
&=& \sum_{\mathbf{q}} \omega_{\mathbf q}\,\hat{a}_{\mathbf{q}}^{\dagger}\,\hat{a}^{\phantom\dagger}_{\mathbf{q}} + \sum_{\mathbf{k}}\,\varepsilon_{\mathbf k}\hat{b}_{\mathbf{k}}^{\dagger}\,\hat{b}^{\phantom\dagger}_{\textbf{k}} \nonumber\\
  &&+ \frac{\lambda}{\sqrt{\mathcal{S}}}\sum_{\mathbf{k},\mathbf{q}}  |\mathbf{q}|\hat{b}_{\mathbf{k}+\mathbf q}^{\dagger}\hat{b}^{\phantom\dagger}_{\mathbf{k}}\,(\hat{a}_{\mathbf{q}}+\hat{a}_{-\mathbf{q}}^{\dagger}),\end{eqnarray}
where $\omega_{\mathbf q}=\hbar^2 q^{2}/2M$ and $\varepsilon_{\mathbf k}= \hbar^2 k^{2}/2m$ are the parabolic dispersion relations   of Tkachenko modes and impurity atoms,  respectively. In the following we set $\hbar = 1$.

\subsection{Perturbative result for polaron Green's function}

In order to calculate the polaron properties, it is natural to use diagrammatic many-body theory. The Green's function of the impurity   can be written as  \be G(\mathbf k,\omega) = [\omega-\varepsilon_{\mathbf k}- \Sigma(\mathbf k,\omega)]^{-1},\ee
where $\varepsilon_{\mathbf k}=k^2/2m$ is the bare impurity dispersion and $\Sigma(\mathbf k,\omega)$ is the self-energy. The polaron energy $E_{\mathbf k}$  is found as the solution to the implicit equation $E_{\mathbf k} \simeq \varepsilon_{\textbf{k}}+\textrm{Re}\Sigma(\mathbf k,E_{\mathbf k})$. For weak interactions, one  usually expects that low energy (small $k$ and $\omega$)  excitations  be quasiparticles that resemble the free particles but with a renormalized mass and a finite lifetime. In this case,
 the retarded  Green's function for frequencies close to the polaron energy can be cast in the form  \cite{mahan}\begin{eqnarray}
G_{ret}(\mathbf k,\omega) \approx \frac{Z}{\omega-  E_{\mathbf k} + i  \gamma_{\mathbf k}}.\label{GwithZ}\end{eqnarray}
Here $Z$ is the  quasiparticle residue (or field renormalization)   given by   \be
Z=\left[1-\left.\left(\frac{\partial{\textrm{Re}\Sigma}}{\partial \omega}\right)\right|_{0}\right]^{-1}.\ee
The decay rate is given by  \begin{equation}
\gamma_{\mathbf k}=-Z \,\textrm{Im}\Sigma_{ret}(\mathbf k, E_{\mathbf k}).\label{gammastar}
\end{equation} Eq. (\ref{GwithZ}) implies that the single-particle spectral function, \be
A(\mathbf k,\omega)=-\frac{1}{\pi}\textrm{Im}G_{ret}(\mathbf k,\omega),\label{defineAkw}
\ee
can be approximated by a Lorentzian peak with weight $Z $ and width $\gamma_{\mathbf k}$.

The expansion of the real part of the self-energy to order $k^2$ yields  a renormalization of the effective mass in the form $E_{\mathbf k}\approx k^2/2m^*$  with
\be
\frac{m}{m^*}=Z \left[1+m\left.\left(\frac{\partial^2\textrm{Re}\Sigma}{\partial k^2}\right)\right|_0\right].\ee
We have omitted a constant energy shift $E_0=\Sigma(\mathbf k=0,\omega=0)$, which we  absorb in the definition of the polaron ground state energy.

\begin{figure}
\begin{center}
\includegraphics*[width=.8\columnwidth]{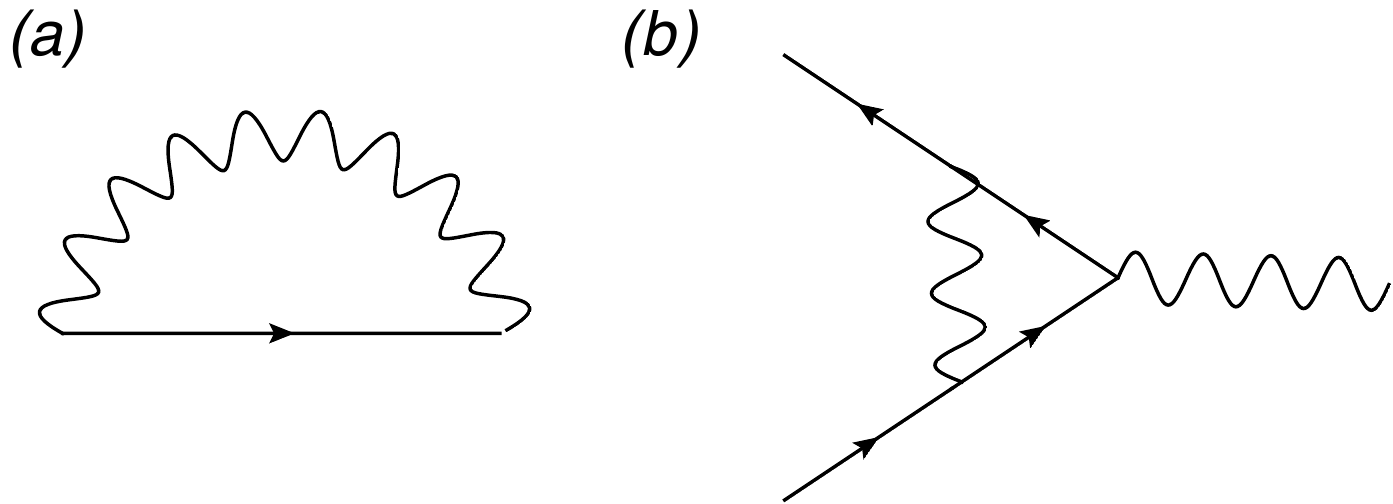}
\label{fig1}
\caption{Impurity self-energy (a) and vertex correction Feynman diagram (b). The solid and wavy lines represent free impurity and Tkachenko mode (analogous to a phonon) propagators, respectively.}
\end{center}
\end{figure}

Let us then  consider the weak coupling limit of Hamiltonian (\ref{eqH}) and calculate   $\Sigma(\mathbf{k},\omega)$ by perturbation theory in $\lambda$. The lowest-order Feynman diagram that contributes to the self-energy contains one Tkachenko mode and  impurity propagator in the intermediate state,    as shown in Fig. 1a. This diagram yields  the retarded self-energy to second order in $\lambda$: \begin{eqnarray} \label{eqtes}
\Sigma^{(2)}_{ret}(\mathbf k,\omega) = \int \frac{d^{2}q}{(2\pi)^{2}}\,\frac{\lambda^2|\mathbf{q}|^{2}}{\omega- \omega_{\mathbf q}-\varepsilon_{\mathbf{k}+\mathbf{q}}+ i \delta},
\end{eqnarray}
where $\delta\to 0^+$. As shown in Ref.  \cite{PRL111},  the decay rate  to order $\lambda^2$ reads\begin{equation}
\gamma_{\mathbf k}\approx -\textrm{Im}\Sigma(k,\varepsilon_{\mathbf k})\approx \frac{\lambda^2mM^3k^2}{2(m+M)^3}.\label{baregamma}
\end{equation}
The perturbative decay rate is linear in energy,  $\gamma_{\mathbf k}\propto \varepsilon_{\mathbf k}\propto k^2$. Therefore, the  quasiparticle peak is only marginally defined, since the relative width  $\gamma_{\mathbf k}/\varepsilon_{\mathbf k}$ does not go to zero as $k\to 0$ (cf. the standard  example of $\mathbf k$ approaching the Fermi surface for quasiparticles  in Fermi liquids \cite{mahan}). Nonetheless, the relative width  can still  be small as long as $\lambda m\ll1$ (assuming $M\gg m$).

The effects of the interaction to order $\lambda^2$ are even more pronounced in the real part of the self-energy. Both the quasiparticle residue and the renormalized mass pick up a    logarithmic dependence on momentum: \begin{equation}\label{Z}
Z(k) \approx \left[1 +  \frac{2\lambda^2\mu^2}{\pi}\;\ln\left(\frac{\Lambda_0}{k}\right)\right]^{-1},
\end{equation}
\begin{equation} \label{mass}
m^*(k)\approx m+\frac{4\lambda^2\mu^3}{\pi}\ln\left(\frac{\Lambda_0}{k}\right),
\end{equation}
where $\Lambda_0$ is an ultraviolet momentum cutoff (of the order of $1/l$) and   $\mu =mM/(m+M)$ is the reduced mass of the two-body problem in the intermediate state (note $\mu \approx m$ for $m\ll M$). Remarkably,  $Z(  k)$  decreases and $m^*( k)$ increases logarithmically as $k$ decreases. Therefore, the perturbative corrections are infrared singular: Even for $\lambda \mu\ll 1$,   the quasiparticle picture  breaks down   at exponentially small momenta   $k\lesssim \Lambda_0 e^{-\pi /(2\lambda^2\mu^2)}$.

Within the perturbative regime $k\gg \Lambda_0 e^{-\pi /(2\lambda^2\mu^2)}$, we may include  the logarithmic corrections in $\gamma_{\mathbf k}$ and $E_{\mathbf k}$ in the spirit of RG improved perturbation theory. We find that  the relative width of the  quasiparticle peak increases logarithmically as $k$ decreases \cite{PRL111} \be
\frac{\gamma_{\mathbf k}}{E_{\mathbf k}}\approx \frac{\lambda^2m ^2}{(1+\varrho)^3}\left[1+\frac{\lambda^2m^2(5+\varrho^2)}{\pi(1+\varrho)^4}\ln\frac{\Lambda_0}{\varepsilon_{\mathbf{k}}}\right],\label{logrelativewidth}
\ee
where $\varrho=m/M $ is the bare  mass ratio.

 \section{Renormalization group analysis}

The logarithmic corrections   appearing  in perturbation theory in Eqs. (\ref{Z}) and (\ref{mass})  can be interpreted in terms of the renormalization of the effective coupling constant $\lambda$ at the  scale set by the impurity momentum $k$. Indeed, in Ref. \cite{PRL111} we derived perturbative RG equations for   $\lambda$ and found it to be  marginally relevant in the weak coupling limit. Here we shall  generalize the RG flow equations to include the renormalization of the quasiparticle weight  and then discuss the low-energy fixed point that arises when $Z(k)\to 0$.

The perturbative RG equations \cite{shankar} can be derived from the one-loop   diagrams in Fig. 1. The self-energy diagram in Fig. 1a and the vertex correction in Fig. 1b are    second order and third order in $\lambda$, respectively. We include the quasiparticle residue $Z(k)$ in the Green's function for the internal impurity lines and consider that the internal momenta are limited by an ultraviolet cutoff $\Lambda$. This cutoff is set by the momentum scale at which we measure correlations of the interacting model, in this case of the order of the impurity momentum, $\Lambda\sim k\ll \Lambda_0$. In the RG step, we consider an infinitesimal reduction of the cutoff to a new value $\Lambda' = \Lambda \, e^{-d\ell}$, with $d\ell \ll 1$, and integrate out fast modes for the impurity and Tkachenko boson with momentum between  $\Lambda^\prime$ and $\Lambda$. Defining the dimensionless parameters  $ \tilde{\lambda} = M \lambda$ and $\tilde m = m/M $, we obtain   the RG equations \begin{eqnarray}\label{RG1}
\frac{dZ }{d\ell} &=& -\frac{\tilde{\lambda}^{2}Z}{\pi}\left(\frac{\mu}{M}\right)^2,\\   \label{RG2}
\frac{d\tilde{\lambda}}{d\ell}&=& \frac{\tilde{\lambda}^{3}Z^{2}}{\pi}\left(\frac{\mu}{M}\right)^2,   \\ \frac{d\tilde{m}}{d\ell} &=& \frac{2\tilde{\lambda}^{2}}{\pi}\;\left(\frac{\mu}{M}\right)^3.\label{RG3}
\end{eqnarray}
The RG flow described by Eqs. (\ref{RG1}), (\ref{RG2}) and (\ref{RG3}) is illustrated in Fig. 2.

 \begin{figure}
\begin{center}
\includegraphics*[width=1.0\columnwidth]{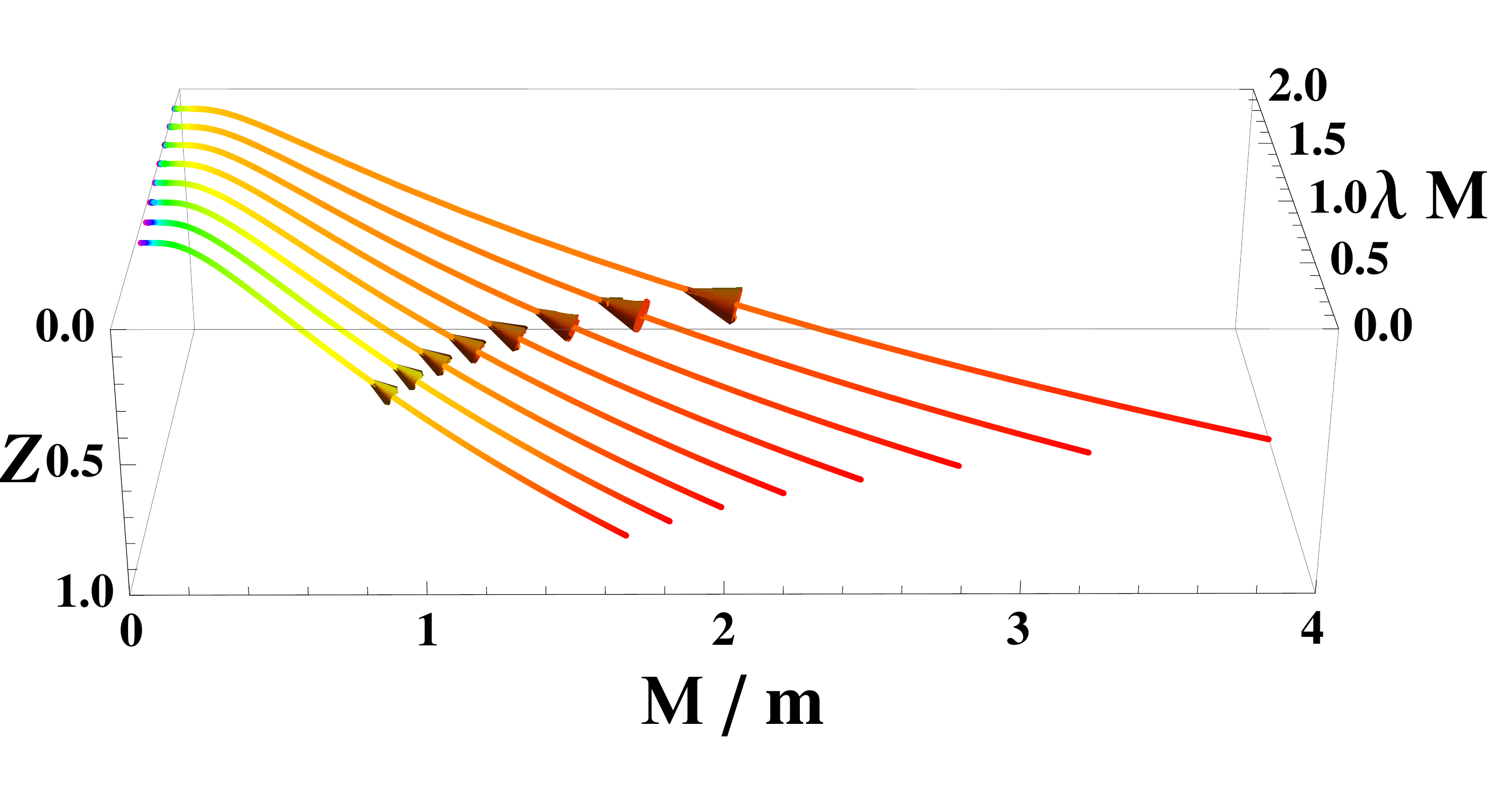}
 \caption{(color online) RG flow diagram.  The    starting points  correspond to  $Z=1$ but different values of the bare dimensionless parameters $\lambda_0 M$ and $M/m_0$. The end points correspond to a line of low-energy fixed points with $Z\to 0$, $M/m\to 0$, but finite $\lambda M$. \label{fig2}}\end{center}
\end{figure}

First, we note that the quasiparticle residue $Z(\Lambda)$ decreases monotonically as we lower the energy scale.  This behavior is reminiscent of another impurity model  studied in condensed matter physics, namely the x-ray edge problem \cite{mahan}. In the latter, the quasiparticle residue of a localized core-hole state that interacts with  low-energy electron-hole pairs in a metal vanishes as a power law in the low energy limit. The power law stems from resumming logarithmic singularities such as the one that appears in Eq. (\ref{Z}), and is a manifestation of the orthogonality catastrophe \cite{noziere,noziere2,noziere3}.  Based on this observation, we expect an analogy between the x-ray edge problem and the physics of the Tkachenko polaron  at low energies  if we replace the core-hole by the mobile impurity and the low-energy electron-hole pairs by Tkachenko modes.

The analogy with the x-ray edge problem can be pursued   further since the solution of the RG equations shows that the effective impurity mass    diverges in the low-energy limit. Thus, there is a crossover from the light impurity   regime, in which $m(\Lambda)\ll M$, to the heavy impurity  regime, in which $m(\Lambda)\gg M$.

Furthermore, we note that the effective coupling constant $\lambda(\Lambda)$ initially grows under the RG flow, in agreement with the results in Ref. \cite{PRL111}. However, the growth is slowed down by the suppression of the quasiparticle weight $Z(\Lambda)$, which affects the vertex correction through the impurity propagators. As a result, for $\Lambda\to0$ the effective $\lambda$ converges to a finite value $\lambda^*$ (see Fig. 2) that depends on the initial value of the bare coupling constant at scale $\Lambda_0$. Therefore, the parameters in the Tkachenko polaron model flow towards a line of fixed points with $Z=0, \tilde m\to\infty$ and continuously varying coupling constant $\lambda^*$. A similar line of fixed points is found in  the Kosterlitz-Thouless flow diagram  which arises for instance in the ferromagnetic regime of the anisotropic Kondo model and   resonant level models \cite{anderson,varma}. Importantly, here the values of $\lambda^*$  are larger than the corresponding bare $\lambda(\Lambda_0)$ only by a factor of order 1. This means that, if we start in the weak coupling  regime $\lambda \mu \ll 1$ with $\mu \approx m(\Lambda_0)$ for a light Tkachenko polaron, the renormalized coupling constant  in the low energy limit may  still be small according to a new criterion $\lambda^*\mu^*\ll 1$, with $\mu^*\approx M$ for $m(\Lambda)\gg M$. In this weak coupling regime, it is justifiable to neglect higher-order corrections in the RG equations.

Close to a fixed point with renormalized coupling constant $\lambda^*$ and in the regime $m(\Lambda)\gg M$, we can simplify the RG equations (\ref{RG1}) and (\ref{RG3}):\bea
\frac{dZ}{d\ell}&\approx &- \frac{(\lambda^*M)^2}{\pi}Z,\\
\frac{d\tilde m}{d\ell}&\approx &\frac{2(\lambda^*M)^2}{\pi}.
\eea
The solution implies that in the low energy limit the quasiparticle weight vanishes as a power law with exponent controlled by the renormalized coupling, $Z(\Lambda)\sim (\Lambda/\Lambda_0)^{ (\lambda^*M)^2/\pi}$, whereas the renormalized mass diverges logarithmically, $\tilde m(\Lambda)\sim \frac{2(\lambda^*M)^2}{\pi}\ln(\Lambda_0/\Lambda)$.

\section{Spectral function in the low energy limit}

The RG analysis in the previous section suggests  a simple picture for the low-energy fixed points of the Tkachenko polaron model in terms of a heavy impurity (with a logarithmically divergent effective mass) with a finite coupling $\lambda^*$  to low-energy Tkachenko modes. In this section we address  the line shape of the single-particle spectral function at low energies, close to a heavy-impurity fixed point.  We first show that, within the approximation of setting $m\to\infty$, the Hamiltonian can be diagonalized exactly and the spectral function is described by a power-law singularity with a nonuniversal   exponent governed by $\lambda^*$. Next, we discuss an approximation to treat the effects of a large but finite impurity mass, the most important of which is to round off the singularity around the renormalized impurity dispersion.

\subsection{Dispersionless impurity}

In the regime $m(\Lambda)\gg M,1/\lambda^*$, we  start with the simplest possible approximation of neglecting the kinetic energy of the impurity in Eq. (\ref{eqH}). In this case of infinite mass, the model is equivalent to a localized impurity coupled to bosonic modes and can be solved exactly by a unitary transformation \begin{eqnarray} \label{eqUnif}
U =  \exp\left[-\frac{1}{\sqrt{\mathcal S}}\sum_{\textbf{q}} \alpha_q\,(\hat{a}_{\textbf{q}}-\hat{a}_{-\textbf{q}}^{\dagger})\hat n_{-\mathbf q}\right],\end{eqnarray}
where $\hat n_{\mathbf q}=\sum_{\mathbf k}\hat{b}_{\textbf{k}}^{\dagger}\hat{b}^{\phantom\dagger}_{\textbf{k}+\textbf{q}}$ is the Fourier transform of the impurity density operator and $\alpha_q$ is a real function of $q=|\mathbf q|$ to be specified below. Eq. (\ref{eqUnif}) is analogous to the Lang-Firsov transformation used in the small polaron regime for lattice models with strong electron-phonon interaction  \cite{mahan}. Using the identity $e^{X}\hat{O}e^{-X} = \hat{O} + [X,\hat{O}]+\frac{1}{2!} [X,[X,\hat{O}]]+ ... $, we obtain the displacement of  the Tkachenko boson  operator    \begin{eqnarray}   \label{displacea}
U^\dagger\hat{a}_{\mathbf{q}}U=\hat{a}_{\mathbf{q}}- \frac{\alpha_{q}}{\sqrt{\mathcal S}}  \hat n_{\mathbf q}. \end{eqnarray}
The transformation of the kinetic energy for Tkachenko modes yields
 \begin{eqnarray} \nonumber
\tilde{H}_{ph} &=& U^\dagger H_{ph} U\nonumber\\
&=& \sum_{\mathbf{q}} \omega_{\mathbf q}\hat{a}^{\dag}_{\mathbf{q}}\hat{a}_{\mathbf{q}} - \frac{1}{\sqrt{\mathcal S}}\sum_{\mathbf{q}}\frac{  \alpha_qq^2}{2M}\hat n_{-\mathbf q}(\hat{a}_{\mathbf{q}}+\hat{a}_{-\mathbf{q}}^{\dag}) \nonumber
\\ &&+\frac1{\mathcal S}\sum_{\mathbf q}\alpha_q^2\frac{q^2}{2M},\label{transfHph}
\end{eqnarray}
where we used $ \hat n_{\mathbf q}\hat n_{-\mathbf q}=\sum_{\mathbf k}\hat b^\dagger_{\mathbf k}\hat b^{\phantom\dagger}_{\mathbf k}=1$ in the subspace with a single impurity.
For the impurity-boson interaction term, we have
\be
\tilde{H}_{imp-ph}=  \frac{\lambda^*}{\sqrt{\mathcal S}}\sum_{\mathbf{q}}q \,\hat n_{-\mathbf q}(\hat{a}_{\mathbf{q}}+\hat{a}_{-\mathbf{q}}^{\dag})  -\frac{2}{\mathcal S}\sum_{\mathbf q}\,\alpha_q\lambda^*q. \label{transfHimpph}\ee
Choosing $\alpha_q=\alpha/q$ with \be\alpha = 2M\lambda^*\label{alphalocalized}\ee
eliminates the impurity-boson coupling in the transformed Hamiltonian:
\begin{eqnarray} \label{eqNewH}
\tilde H=U^{\dagger} H U = \sum_{\textbf{q}} \frac{q^{2}}{2M}\,\hat{a}_{\textbf{q}}^{\dagger}\,\hat{a}_{\textbf{q}}+ const.\end{eqnarray}

After the unitary transformation, the Hamiltonian is noninteracting and its ground state  is a vacuum of the transformed bosonic modes in Eq. (\ref{eqNewH}).  The ground state $|\tilde 0\rangle$ in the presence of the impurity  is related to the original bosonic vacuum $|0\rangle$  by $ | \tilde{0} \rangle=U  |0\rangle$. Thus,  the overlap  between the ground states with or without the  impurity   is   \bea \label{eq1}
\langle 0 | \tilde{0} \rangle&=&\langle 0 | \exp\left[-\frac{\alpha}{\sqrt{\mathcal S}}\sum_{\mathbf{q}} \frac{1}{q}\,(\hat{a}_{\textbf{q}}-\hat{a}_{-\textbf{q}}^{\dagger})\right] | 0 \rangle\nonumber\\
& = &e^{-\frac{\alpha^2}{\mathcal S}\sum_{\mathbf{q}} \frac{1}{q^2}}. \quad \eea
For a large vortex lattice, the sum can be converted into an integral $\frac{\alpha^2}{\mathcal S}\sum_{\mathbf{q}} \frac{1}{q^2} \rightarrow \frac{\alpha^2}{2\pi}  \int_{q_{min}}^{\Lambda_0} \frac{dq}{q}$, which  diverges logarithmically for $q_{min}\to0$. We   cut off the infrared divergence by setting the lower limit of integration to be $q_{min}\sim 2\pi/L$ with $L\sim \sqrt{\mathcal S}$ the length scale representing the system size. We then find \be
\langle 0 | \tilde{0} \rangle\sim L^{-\alpha^2/2\pi}.\label{OCoverlap}\ee
Thus,  the overlap vanishes in the thermodynamic limit. As usual, the orthogonality catastrophe stems from the creation of a divergent number of low-energy, small-$q$ excitations --- in this case the  Tkachenko bosons ---  upon coupling the many-body system  to a single impurity.

We can use the unitary transformation to calculate the impurity Green's function   \be
G(\mathbf r,t>0)= -i \langle  \hat\psi_B(\mathbf{r},t)\hat\psi_B^{\dagger}(\mathbf{0},0)\rangle ,
\ee
where $\langle\,\rangle$ denotes the expectation value in the ground state of the system without    impurities.  Using  the mode expansion of the impurity field operator $ \hat\psi_B(\mathbf{r},t)=\frac{1}{\sqrt{\mathcal S}}\sum_{\mathbf{k}} e^{i\mathbf{k}.\mathbf{r}}\,\hat{b}_{\textbf{k}}(t)$, it is easy to show that \begin{eqnarray}  \nonumber \tilde{\psi}_B(\mathbf{r},t) &=& U^\dagger\hat\psi_B(\mathbf{r},t) U=\hat\psi_B(\mathbf{r},t) \;e^{\alpha Y(\mathbf{r},t)} , \end{eqnarray} where $Y(\mathbf r)$ is the anti-hermitean  displacement operator for Tkachenko modes\be
Y(\mathbf{r})=  \frac{1}{\sqrt{\mathcal S}}\sum_{\mathbf{q}} \frac{e^{i\mathbf{q}\cdot\mathbf{r}}}{q}(\hat{a}_{\textbf{q}}-\hat{a}_{-\textbf{q}}^{\dagger}).\ee
The impurity Green's function becomes \be  \label{green1}
G(\mathbf{r},t) =- i  \langle \tilde\psi_B(\mathbf{r},t)\tilde\psi_B^{\dagger}(\mathbf{0},0)\rangle  \langle e^{-\alpha Y(\mathbf{r},t)}e^{\alpha Y(\mathbf{0},0)}\rangle, \ee
where we used the decoupling of impurity and Tkachenko modes in the transformed Hamiltonian, with $\langle \, \rangle$ the free impurity background.
The  field $\tilde\psi_B^\dagger$ creates a free impurity with infinite mass at position $\mathbf r$. The operator  $e^{-\alpha Y(\mathbf r)}$ can be interpreted as creating the cloud of Tkachenko bosons around the impurity.  Since the problem is now noninteracting, we can calculate the exact propagators. For the impurity term we have \begin{eqnarray}
\label{eq1} \langle\tilde\psi_B(\mathbf{r},t)\tilde\psi_B^{\dagger}(\mathbf{0},0)\rangle = \frac{1}{\mathcal S}\sum_{\textbf{k}}e^{i\mathbf{k}\cdot\mathbf{r}}=\delta(\mathbf r), \end{eqnarray}
where we used $\varepsilon_{\mathbf k}=0$ for $m\to\infty$. For the bosonic part in Eq. (\ref{green1}) we use the Baker-Hausdorff formula $e^{A+B} = e^{A} e^{B} e^{-[A,B]/2}$ to rewrite the operators in normal order,  and obtain
\bea
\langle e^{-\alpha Y(\mathbf{r},t)}e^{ \alpha Y(\mathbf{0},0)}\rangle &=& e^{\alpha^{2}I(\mathbf{r},t)} ,\label{propagY}
\eea
with  $I(\mathbf{r},t) = \frac{1}{2\pi} \int\frac{dq}{q}\left[\exp(-i \omega_{q} t) J_0(qr)-1 \right]$, where   $J_0(x)$ is the Bessel function of the first kind. Due to the delta function at the position of the impurity  in Eq. (\ref{eq1}), we can set $\mathbf r=0$ in Eq. (\ref{propagY}). For $t\gg (\Lambda_0^2/2M)^{-1}$, we have\be
I(\mathbf{r}=0,t) \approx -\frac{1}{4\pi}\left[\gamma + \ln\left( \frac{i \Lambda_0^2  t}{2M}\right)\right],\label{Irt}\ee
where $\gamma$ is  Euler's constant. Substituting Eq. (\ref{Irt}) in Eqs. (\ref{propagY}),  we obtain a power-law decay \be
\langle e^{-\alpha Y(\mathbf{0},t)}e^{ \alpha Y(\mathbf{0},0)}\rangle\propto t^{-\alpha^{2}/4\pi}.\label{powerlawdecay}
\ee
The Green's function for the localized impurity is then \begin{eqnarray}\label{green}
G(\mathbf{r},t) \propto \delta(\mathbf{r})\;t^{-\alpha^{2}/4\pi}.
\end{eqnarray}
The nonuniversal exponent $\alpha^{2}/4\pi = (M\lambda^*)^{2}/\pi $ is consistent with the result for the orthogonality catastrophe in Eq. (\ref{OCoverlap}). The spectral function defined in Eq. (\ref{defineAkw}) can be calculated by taking the Fourier transform of Eq. (\ref{green}).  The result is a  power-law singularity
\begin{eqnarray}
A(\mathbf{k},\omega) \sim \omega^{-1+\alpha^{2}/4\pi}.
\end{eqnarray}
We obtain a divergent singularity if the renormalized coupling obeys the condition $(M\lambda^*)^2< \pi$, which is verified in the perturbative regime  $M\lambda^*\ll1$.

\subsection{Finite impurity mass}

When the impurity mass $m$ is finite, the unitary transformation in Eq. (\ref{eqUnif}) does not diagonalize the Hamiltonian exactly because the transformation of the impurity kinetic energy  $H_{imp}$  generates  additional interactions:\bea \label{eqHT}
 \tilde{H}_{imp} &=& \sum_{\mathbf{k}} \varepsilon_{\mathbf k}\hat{b}^{\dag}_{\mathbf{k}}\hat{b}^{\phantom\dagger}_{\mathbf{k}} +\frac1{\sqrt{\mathcal S}}\sum_{\mathbf q}\alpha_q \hat{\mathbf J}_{-\mathbf{q}}\cdot \mathbf q (\hat{a}_{\mathbf q}-\hat{a}_{-\mathbf{q}}^{\dag})\nonumber\\
&&+\frac1{2\mathcal S}\sum_{\mathbf q,\mathbf q^\prime} \alpha_q \alpha_{q^\prime}\frac{\mathbf q\cdot \mathbf q^\prime}{m}\times\nonumber\\
&&\times\hat n_{-\mathbf q-\mathbf q^\prime}(\hat{a}^{\phantom\dagger}_{\mathbf{q}}-\hat{a}_{-\mathbf{q}}^{\dag})(\hat{a}^{\phantom\dagger}_{\mathbf{q}^\prime}-\hat{a}_{-\mathbf{q}^\prime}^{\dag}),
\eea
where \be
\hat {\mathbf J}_{\mathbf q}=\frac1m\sum_{\mathbf k}\left(\mathbf k+\frac{\mathbf q}2\right)\hat{b}^{\dag}_{\mathbf{k}}\hat{b}^{\phantom\dagger}_{\mathbf{k}+\mathbf{q}}\ee
is the impurity current operator.

Here we resort to a variational method based on a partial polaron transformation \cite{siebel,agarwal}. The idea is that, when the impurity mass is large compared to the Tkachenko boson mass, the bosonic field can instantaneously adjust to the slow motion of the  impurity. We then employ a variational ground state which is a vacuum of bosons in the appropriate representation. In practice, we  perform  a unitary transformation of the form in Eq. (\ref{eqUnif}) and fix the function $\alpha_q$ by the condition that the polaron ground state energy be minimized in a new boson vacuum  $|\tilde 0\rangle$ in the presence of the impurity. Taking the expectation value of Eqs. (\ref{eqHT}), (\ref{transfHph}) and (\ref{transfHimpph}) in a  state with the impurity with momentum $\mathbf k$,   we obtain the energy as a functional of $\alpha_q$ \bea
 \mathcal{E}_\mathbf{k}(\alpha_q)&=& \frac{k^2}{2m} +  \frac{1}{\mathcal{S}} \sum_{\mathbf{q}}\left( \frac{\alpha_{q}^2q^2}{2\mu}-\frac{2 \lambda^* \alpha_{q} }{q} \right).\label{functional}\eea
Minimizing the energy in Eq. (\ref{functional}) with respect to $\alpha_q$, we find\be
\alpha_q=2\mu \lambda^*/q,\label{newalpha}
\ee
which differs from the result in Eq. (\ref{alphalocalized}) for an infinite-mass impurity only in that the Tkachenko boson mass is replaced by the reduced mass $\mu$.

After the unitary transformation with $\alpha_q$ given in Eq. (\ref{newalpha}), the transformed Hamiltonian is still interacting,\begin{eqnarray} \nonumber \label{eq21} \tilde{H} &=& \sum_{\textbf{q}} \omega_{\mathbf q}\hat{a}_{\textbf{q}}^{\dagger}\hat{a}^{\phantom\dagger}_{\textbf{q}} + \sum_{\textbf{k}} \varepsilon_{\mathbf k}\hat{b}_{\textbf{k}}^{\dagger}\hat{b}^{\phantom\dagger}_{\textbf{k}} +H_r,
\eea
with the residual interactions\bea
H_r  &  =&  \frac{\lambda^\prime}{\sqrt{\mathcal S}}\sum_{\textbf{k},\textbf{q}}   q\,  \hat{b}_{\textbf{k}+\textbf{q}}^{\dagger}\hat{b}^{\phantom\dagger}_{\textbf{k}}\,(\hat{a}_{\textbf{q}}+\hat{a}_{-\textbf{q}}^{\dagger})\nonumber
  \\    & &+\frac{2\lambda^\prime}{\sqrt{\mathcal S}}\sum_{\textbf{k},\textbf{q}}  \frac{\mathbf q}{q}\cdot\left(\mathbf k -\frac{\mathbf q}2\right) \hat{b}_{\textbf{k}}^{\dagger}\hat{b}^{\phantom\dagger}_{\textbf{k}-\textbf{q}}\,(\hat{a}_{\textbf{q}}-\hat{a}_{-\textbf{q}}^{\dagger}).\label{residual} \end{eqnarray}
Importantly, the residual interaction involves   a rescaled  coupling constant $\lambda^\prime=\lambda^*M/(m+M)$, which is  suppressed by the renormalized mass ratio $M/m\ll 1$ at low energies. In Eq. (\ref{eq21}) we have  discarded    terms of   order  $(M\lambda^*)^2$ in the perturbative regime $M\lambda^*\ll 1$.

If  we first neglect the residual interactions of order $\lambda^\prime\ll \lambda^*$ in Eq. (\ref{eq21}),  the Green's function  still factorizes as in  Eq. (\ref{green1}). The only difference is that for a finite mass $m$ the free impurity propagator becomes \be
\langle {\tilde\psi_B}(\mathbf{r},t){\tilde\psi_B}^{\dagger}(\mathbf{0},0)\rangle= \frac{m}{2\pi i t} \,e^{imr^{2}/2t}.\label{propwithmass}\ee
The Fourier transform of the Green's function  is
\be
G(\mathbf{k},\omega) = -\int d^2r dt \,  e^{- i \mathbf{k}\cdot \mathbf{r}+i\omega t}\,   \frac{m e^{imr^{2}/2t}}{2\pi t} e^{\bar \alpha^2 I(\mathbf r,t)},\label{fourierG}
\ee
where $\bar \alpha =2\mu \lambda^*$.
For $m\gg M$, the fast spatial oscillations in the propagator  (\ref{propwithmass}) imply that the integral in Eq. (\ref{fourierG}) is dominated by short distances $r  \lesssim \sqrt{t/m}$. This allows us to neglect the spatial dependence of $I(\mathbf{r},t)$. Physically, this means the Tkachenko mode diffuses much faster than the impurity and  the dominant contribution  stems from  the long-time tail of $e^{\bar\alpha^{2}I(\mathbf{r},t)}$ near the origin.  We then  approximate \bea
G(\mathbf{k},\omega) &\approx&  \int dt \,e^{i\omega t}e^{\bar \alpha^2 I(\mathbf 0,t)}  \int d^2r e^{- i \mathbf{k}\cdot \mathbf{r}}\,   \frac{m}{2\pi i t} \,e^{imr^{2}/2t}\nonumber\\
&=& \int dt \,e^{i(\omega-\varepsilon_{\mathbf k}) t}e^{\bar\alpha^2 I(\mathbf 0,t)}.\label{Gnewthreshold}\eea
Eq. (\ref{Gnewthreshold}) involves the Fourier transform of the power-law decaying boson cloud propagator in Eq. (\ref{powerlawdecay}). However, the frequency  dependence is shifted, and the spectral function develops a power-law singularity above the single-impurity threshold \be
A(\mathbf k,\omega) \sim \theta(\omega-\varepsilon_{\mathbf k})(\omega-\varepsilon_{\mathbf k})^{-1+\eta},\label{approxAnores}
\ee
where $\eta =\bar \alpha^2/4\pi$ and $\theta(x)$ is the Heaviside step function.

The vanishing of the spectral function for $\omega<\varepsilon_{\mathbf k}$ in Eq. (\ref{approxAnores}) is an artifact of neglecting the residual interactions. In fact, kinematics implies that the support of the spectral function for any $\mathbf k$ must extend to arbitrarily low energies, since the impurity can always decay  by emitting   bosons with parabolic dispersion \cite{PRL111}. This also means that the approximate power law singularity obtained in Eq. (\ref{approxAnores}) is inside a multiparticle  continuum and must be broadened when we take into account the impurity decay due to the residual interactions.

 \begin{figure}
\begin{center}
\includegraphics*[width=1.0\columnwidth]{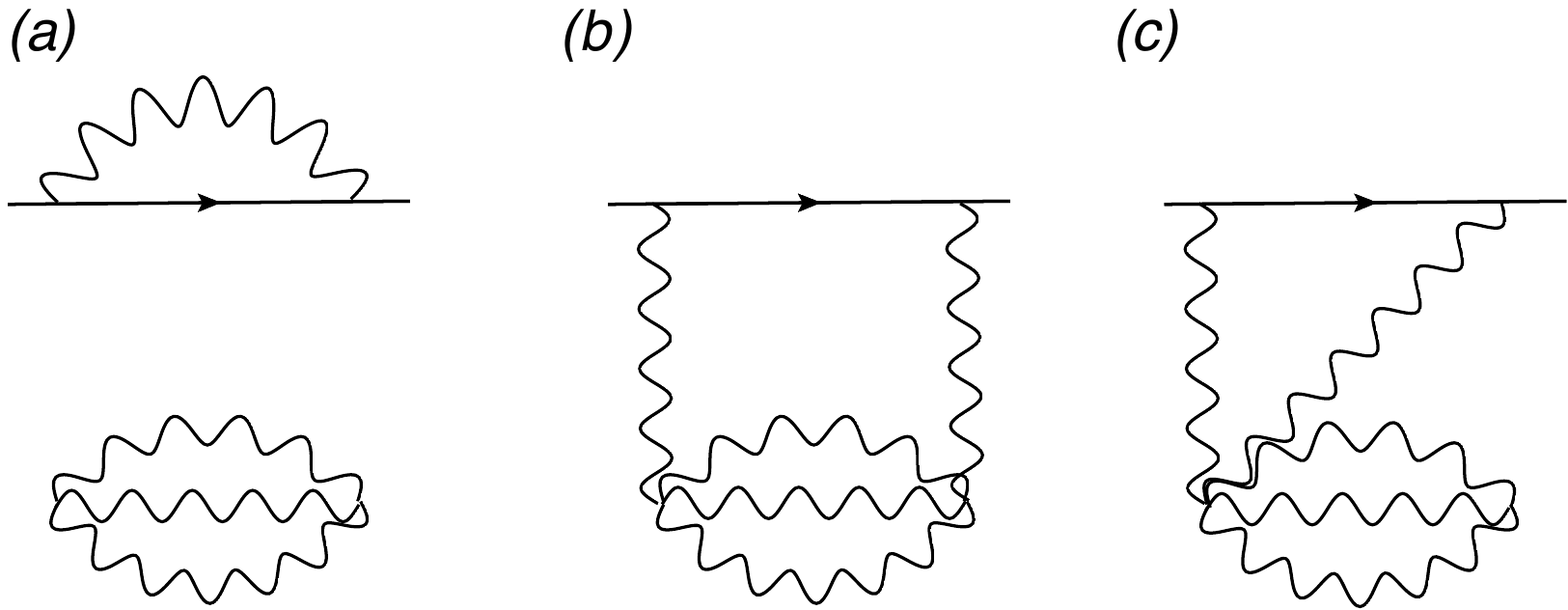}
\caption{(a) Lowest order diagram included in the approximation for  the impurity Green's function in the limit $m\gg M$. The disconnected part with multiple wiggly lines  represents the boson cloud propagator. (b) and (c) Connected cloud diagrams neglected in the approximation. \label{fig3}}
\end{center}
\end{figure}

To obtain the rounding of the singularity for  $m/M\ll 1$, we apply perturbation theory in interactions \eqref{residual} to calculate the Green's function $G(\mathbf{r},t) = -i \langle \tilde\psi_B(\mathbf{r},t)\tilde\psi_B^{\dagger}(\mathbf{0},0)  e^{-\bar\alpha Y(\mathbf{r},t)}e^{\bar\alpha Y(\mathbf{0},0)}e^{-i\int dt' H_{r}(t')}\rangle_0$.
 To order $(\lambda^\prime)^2$, the simplest diagram has a self-energy insertion in the impurity propagator  (see Fig. \ref{fig3}). There are, in addition, diagrams in which the impurity propagator is connected with the cloud propagator by taking contractions of the bosonic operators in $H_r$ with the operators $e^{\bar \alpha Y(\mathbf r)}$. However, the latter  type of contraction introduces an additional factor of $\bar \alpha^2\sim (\mu \lambda^*)^2$. Thus, in the perturbative regime $M\lambda^*\ll 1$ we   neglect  diagrams that connect the  impurity to the boson cloud propagator and sum up the series of self-energy diagrams such as the one in Fig. \ref{fig3}a. Within this approximation, we have $G(\mathbf{r},t) = \mathcal{G}(\mathbf{r},t)  e^{\bar \alpha^2 I(\mathbf{r},t)}$, where $\mathcal{G}(\mathbf{r},t)$ is the  impurity Green's function dressed with the self-energy from the residual interactions. Taking the Fourier transform of $G(\mathbf r,t)$ and neglecting the spatial dependence of $I(\mathbf r,t)$, we obtain  \be \label{eqG}
G(\mathbf{k},\omega) \approx  \frac{2e^{-\gamma \eta}\sin(\pi \eta)\Gamma(1-\eta)}{(i\Lambda_0^2/2M)^{\eta}} \int_{\omega}^{\infty} \frac{d\omega'}{2\pi}\frac{ \mathcal{G}(\mathbf{k},\omega')  }{(\omega'-\omega)^{1-\eta}}.
\ee
The dressed impurity Green's function is\be
\mathcal{G}(\mathbf{k},\omega) =\frac{1}{\omega-\varepsilon_{\mathbf{k}}-\Sigma_r(\mathbf{k},\omega)}.\ee

The real part of $\Sigma_r(\mathbf k,\omega)$ is cutoff dependent and contains logarithmic divergences, which appear again because the residual interaction is marginal. These divergences must be absorbed in the definition of the renormalized parameters. We are mainly interested in the imaginary part
\begin{eqnarray}
\mathrm{Im}\Sigma_r(\mathbf{k},\omega) &=& - (\lambda^\prime)^2 \left[ \mu k^2+2m^2(\varepsilon_{\mathbf k}-\omega)\theta(\varepsilon_{\mathbf k}-\omega)\right]\times \nonumber  \\&& \times \theta[\omega-\varepsilon_{min}(k)],  \end{eqnarray}
where $\varepsilon_{min}(k)=k^2/[2(m+M)]$ is the lower threshold of the two-particle (impurity plus one boson) continuum. The threshold at $\omega=\varepsilon_{min}(k)$ is an artifact of calculating $\Sigma_r$ only to order $(\lambda^\prime)^2$. At higher orders in perturbation theory $\textrm{Im}\Sigma(\mathbf k,\omega)$ must be nonzero  for any  $\omega>0$. But notice that the energy window between the two-particle lower threshold  and the single-impurity energy, $\delta \varepsilon_k=\varepsilon_{\mathbf k}-\varepsilon_{min}(k)\approx Mk^2/m^2$, vanishes more rapidly than $\varepsilon_{\mathbf k}$ as the effective mass $m(k)$ diverges for $k\to0$. This is expected since the phase space available for scattering decreases as the impurity becomes heavier and the recoil energy vanishes \cite{noziere}. The same behavior is observed for the decay rate calculated from the residual interaction\be
\gamma^{r}_{\mathbf k}\approx -\textrm{Im}\Sigma_r(\mathbf k,\omega=\varepsilon_{\mathbf k}) =(\lambda^\prime)^2\mu k^2,
\ee
which also scales like $\sim 1/m^2$ for $m\gg M$. 

\begin{figure}
 \includegraphics*[width=1.0\columnwidth]{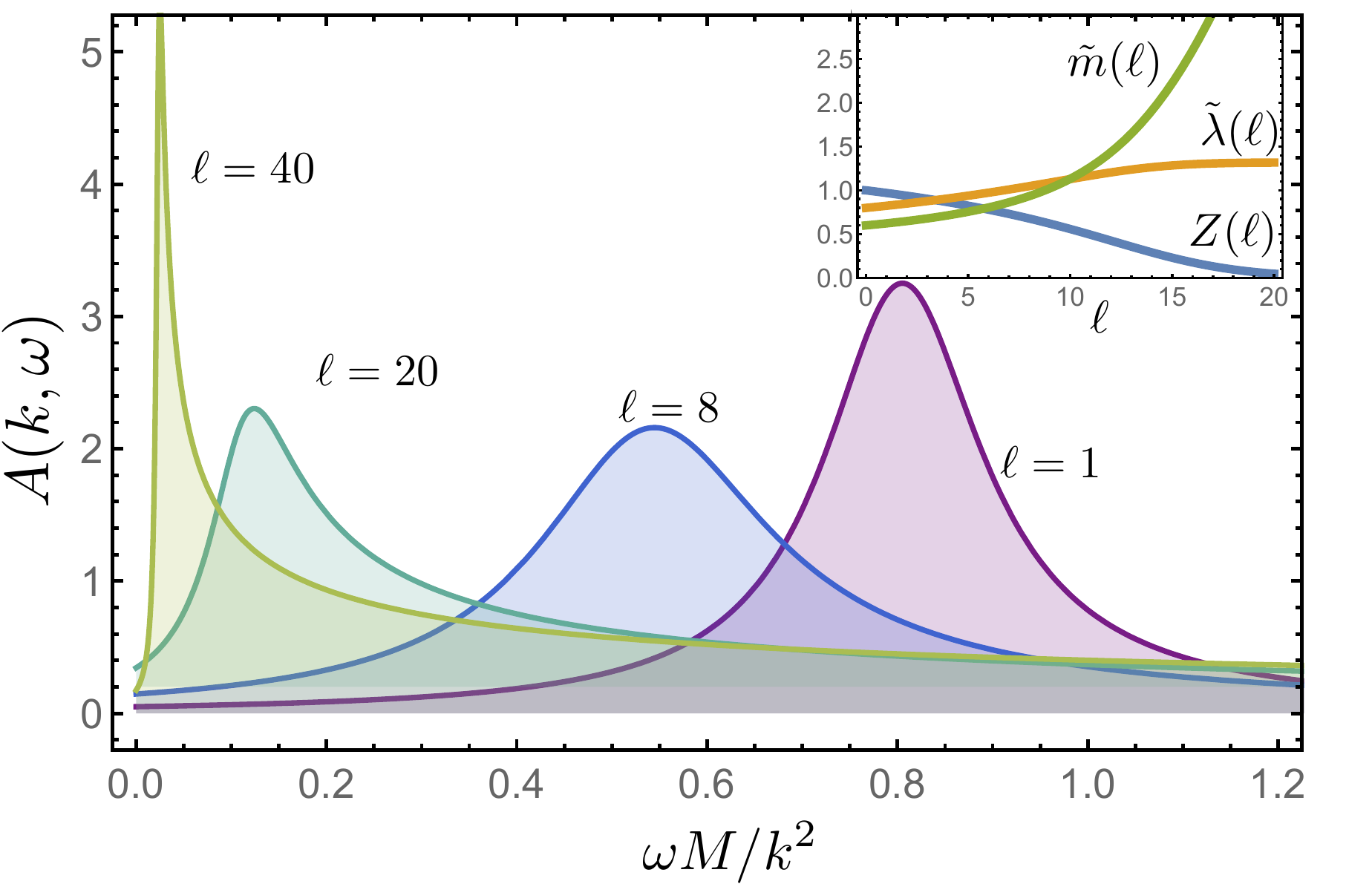}
\caption{(color online) Impurity spectral function $A( k,\omega)$ in the low energy regime for four different values of $k=\Lambda_0e^{-\ell}$.  As $\ell$ increases, the line shape crosses over from a Lorentzian to an approximate power-law singularity. The inset shows the RG flow  of the dimensionless parameters; here we set the bare values to $Z(0)=1,\tilde m(0)=0.6, \tilde \lambda(0)=0.8$. Here we set $\Lambda_0=M=1$. The spectral function is normalized  so as to obey the sum rule \cite{mahan} $\int_0^{K} d\omega\,  A(k,\omega)=1$, with a  cutoff on the high-frequency tail  of Eq. (\ref{spectral}) set by $K=k^2/[M\tilde m(0)]$. \label{fig4}  }
\end{figure}

Since we are interested in the broadening of the spectral function in a small energy window $|\omega-\varepsilon_{\mathbf k}|/\varepsilon_{\mathbf k}\sim M/m\ll 1$, we approximate $\Sigma_r(\mathbf k,\omega)\approx \Sigma_r(\mathbf k,\varepsilon_{\mathbf k})$. We absorb the real part of the self-energy into the renormalized dispersion $E_{\mathbf k}$ and write the imaginary part as the decay rate $\gamma^r_{\mathbf k}$. Then from Eq. (\ref{eqG}) we obtain the  result for the spectral function \be
\label{spectral} A(\mathbf{k},\omega)   \propto  \frac{\sin\left\{(1-\eta) \left[\frac{\pi}{2}-\arctan\left(\frac{E_{\mathbf{k}}-\omega}{\gamma^r_{\mathbf{k}}}\right)\right]\right\}}{\left[(\omega-E_{\mathbf{k}})^2+(\gamma^r_{\mathbf{k}})^{2}\right]^{(1-\eta)/2} }.\ee
The spectral function has a broadened peak at $\omega\approx E_{\mathbf k}$; but since  $\gamma^r_{\mathbf{k}}/E_{\mathbf{k}}$ decreases as the effective impurity mass increases,   the peak      becomes more pronounced for smaller $k$ and we recover the power-law singularity in the limit $k \rightarrow 0$, $m \rightarrow \infty$. The change in the line shape of $A(\mathbf k,\omega)$ between the large $k$ and small $k$ regimes is illustrated in Fig. \ref{fig4}.  

\section{Probing the excitation spectrum}


Analogous to angle resolved photoemission spectroscopy (ARPES) used to measure  the single-particle spectral function  of  electrons in metals, in   cold atom setups  there is the technique of  momentum resolved radio-frequency (rf) spectroscopy \cite{T1,T2}. Basically, a  rf  light pulse is used to transfer   impurity atoms to a hyperfine level that does not interact with the background atoms. After that, a free expansion absorption image provides the momentum distribution of the impurity sample. Since the rf pulse does not alter the original atomic momentum, one can recover the impurity single-particle spectrum  combining the information from  the hyperfine level separation (energy of the light pulse) and the release energy of the free expanding atoms. To probe different regimes in the spectral function, we should start with an external force that acts selectively on impurity atoms (through a magnetic field gradient \cite{shashi} or  a two-photon stimulated Raman
transition   \cite{ketterle, berciu}) to impart a well-defined initial momentum. Then a rf pulse can be applied, after an appropriate time interval, to transfer the initially interacting impurities to a noninteracting final state. The release energy of the dilute impurity sample can be measured trough the time of flight state-selective absorption image, realized with the same holding time, but for different impurity momenta applied initially.

As in the x-ray edge problem \cite{mahan}, the single-particle Green's function in real time $G(\mathbf k,t)$ can be related to a time-dependent  overlap $ \langle   0_k |  e^{iH_At}e^{-i (H_A+H_B+H_{int}) t} |   0_k \rangle$, where $|\tilde 0_k\rangle =\hat b^\dagger_{\mathbf k}|0\rangle $. It has been proposed that this type of overlap can be measured directly  using Ramsey-type interferometry \cite{PRX}. In our case, measuring the decay of the overlap with time would be useful to distinguish between the two regimes in the spectral function. For momentum $l^{-1}e^{-\pi/4\mu_{0}^{2}\lambda_{0}^{2}} \ll k\ll l^{-1}$, we expect an exponential decay $\sim e^{-\gamma_{\mathbf k}t} $ controlled by the width $\gamma_{\mathbf k}$ of the Lorentzian peak in $A(\mathbf k,\omega)$. In the long wavelength  regime $k\ll l^{-1}e^{-\pi/4\mu_{0}^{2}\lambda_{0}^{2}}$ and for intermediate times $(Ml^2)^{-1}\ll t\ll (\gamma^r_{\mathbf k})^{-1}$, one should observe a power law decay $\sim t^{-\eta}$ as a signature of the orthogonality catastrophe and breakdown of the quasiparticle  picture for  the Tkachenko polaron.

While we have emphasized the crossover in the spectral function as a function of momentum, the orthogonality catastrophe in the vortex lattice could also be observed using impurities localized by an external potential. In this case it would suffice to measure the frequency dependence in rf spectroscopy, removing the need for momentum resolved techniques. 

\section{Conclusion}

We have studied the model  of a neutral impurity weakly coupled with the Tkachenko modes of a vortex lattice Bose-Einstein condensate. We have described how the  line shape of  the impurity spectral function is modified as the impurity momentum varies between a perturbative regime $k\gg l^{-1}e^{-\pi/4\mu_{0}^{2}\lambda_{0}^{2}}$ and a low energy regime $k\ll l^{-1}e^{-\pi/4\mu_{0}^{2}\lambda_{0}^{2}}$.  In the low energy limit the spectral function develops  a  power law singularity. The latter is a signature of the orthogonality catastrophe that arises as the effective  impurity mass $m(k)$ grows with the RG flow and the heavy impurity is dressed by an increasing  number of low-energy Tkachenko modes. For any  $k>0$, the singularity is broadened due to the recoil of the finite mass impurity, but the singularity becomes well defined  in the limit $k\to0$. We have proposed that the crossover in the line shape of the Tkachenko polaron spectral function could be measured using momentum-resolved radio-frequency spectroscopy. 
\vspace{0.9 cm}
\acknowledgements

This work is supported by Fapesp/CEPID (M.A.C.) and CNPq (R.G.P.).

\end{document}